\documentclass[aoas]{imsart}

\RequirePackage{amsthm,amsmath,amsfonts,amssymb}

\RequirePackage[authoryear]{natbib}

\RequirePackage[colorlinks,citecolor=blue,urlcolor=blue]{hyperref}
\RequirePackage{graphicx}
\usepackage{subcaption} 
\usepackage{algorithm}
\usepackage{algpseudocode}

\usepackage{color}

\startlocaldefs
\theoremstyle{plain}

\newtheorem{theorem}{Theorem}[section]

\newtheorem{proposition}{Proposition}[section]

\theoremstyle{remark}

\endlocaldefs

\begin{document}

\begin{frontmatter}
\title{Weak Signal Inclusion Under Dependence and Applications in Genome-wide Association Study}
\runtitle{Weak Signal Inclusion}

\begin{aug}
\author[A]{\fnms{X. Jessie} \snm{Jeng}\ead[label=e1,mark]{xjjeng@ncsu.edu}},
\author[B]{\fnms{Yifei} 
\snm{Hu}\ead[label=e2]{hyifei@ncsu.edu}},
\author[C]{\fnms{Quan} 
	\snm{Sun}\ead[label=e3]{quansun@live.unc.edu}}
\and
\author[D]{\fnms{Yun} 
	\snm{Li}\ead[label=e4]{yunli@med.unc.edu}}

\address[A]{Department of Statistics, North Carolina State University, \printead{e1}}

\address[B]{Department of Statistics, North Carolina State University, \printead{e2}}

\address[C]{Department of Biostatistics, University of North Carolina at Chapel Hill, \printead{e3}}

\address[D]{Department of Biostatistics, Genetics, University of North Carolina at Chapel Hill, \printead{e4}}

\end{aug}

\begin{abstract}
In this study, we present a data-driven method called false negative control (FNC) screening to address the challenge of detecting weak signals in underpowered genome-wide association studies (GWASs), where true signals are often obscured by a large amount of noise. Our approach focuses on controlling false negatives and efficiently regulates the proportion of false negatives at a user-specified level in realistic settings with arbitrary covariance dependence between variables. We calibrate overall dependence using a parameter that aligns with the existing phase diagram in high-dimensional sparse inference, allowing us to asymptotically explicate the joint effect of covariance dependence, signal sparsity, and signal intensity on the proposed method. Our new phase diagram shows that FNC screening can efficiently select a set of candidate variables to retain a high proportion of signals, even when the signals are not individually separable from noise. We compare the performance of FNC screening to several existing methods in simulation studies, and the proposed method outperforms the others in adapting to a user-specified false negative control level. Moreover, we apply FNC screening to 145 GWAS datasets obtained from the UK Biobank and demonstrate a substantial increase in power to retain true signals for downstream analyses. 
\end{abstract}


\end{frontmatter}

\section{Introduction}  \label{sec:introduction}

Genome-wide association studies (GWAS) have successfully identified genetic variants that predispose individuals to complex traits and diseases since their establishment in the early 2000s, often by utilizing very large sample sizes \citep{Hu2021RBC, Mikhaylova2021WBC, Zhao2022NG, Mahajan2022T2D}. However, GWAS of many diseases remain underpowered, particularly for non-European ancestries \citep{Sun2022UKBminority, Liu2022ELGAN}. This issue is further compounded in the field of integrative genomics, where the analysis of multi-omics data is impeded by the limited availability of smaller sample sizes compared to GWAS. It is now understood that the bulk of the missing heritability can be attributed to a large number of common variants with small effect sizes. Statistical methods have not adequately addressed such weak signals. 

\begin{figure}[tb]
	\centering
	\includegraphics[height = 5cm, width=6.5cm]{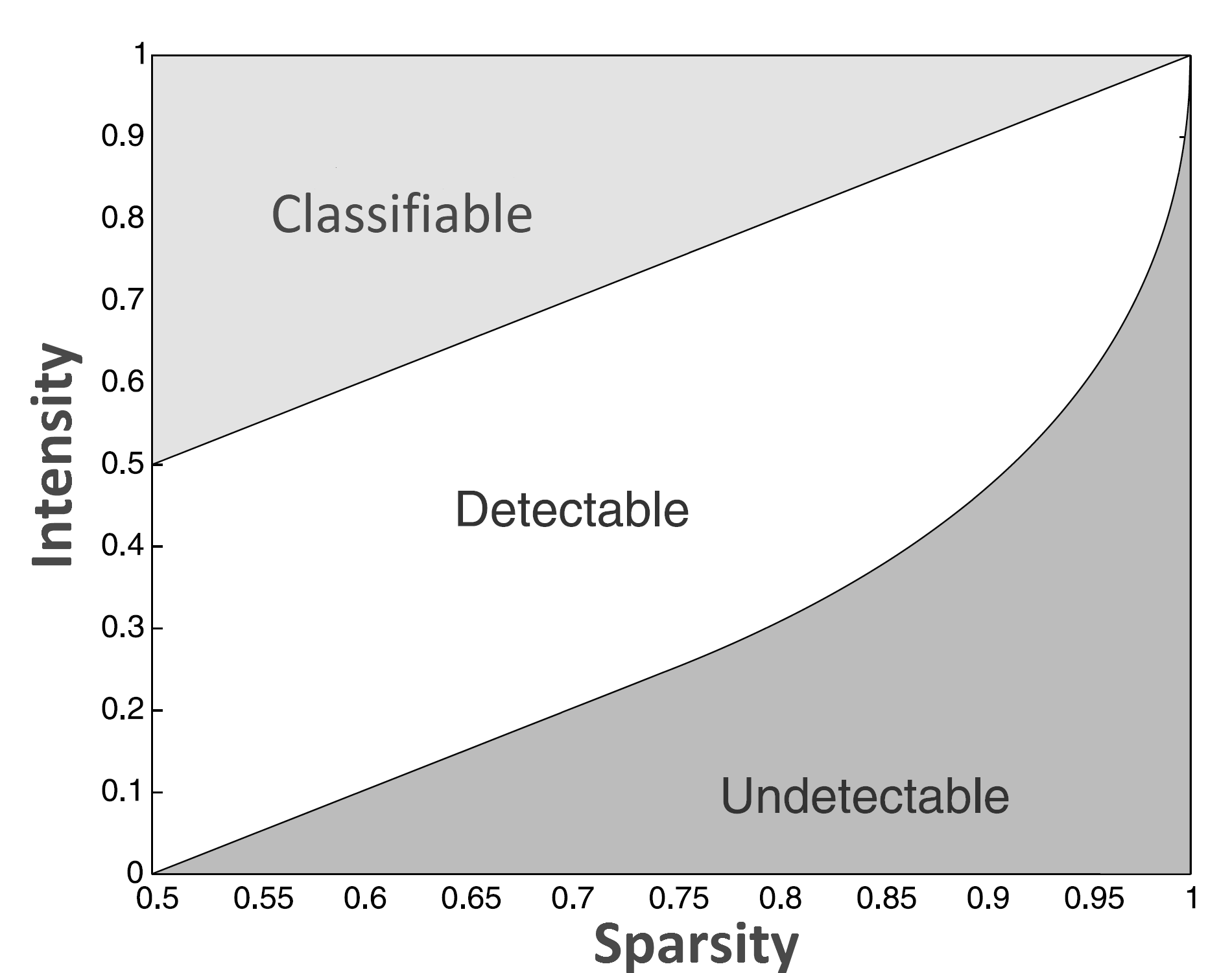}
	\caption{Phase diagram for signal detection
		and classification. Signals in the middle region are only detectable for their existence but cannot be well-separated from noise.} \label{fig:phase}
\end{figure}

Over the years, significant contributions have been made towards two related problems in high-dimensional statistical inference. The first problem is the detection of mixture models, which involves detecting the existence of sparse signals without specifying their exact locations (see, e.g., \cite{donoho2004higher}, \cite{arias2011global}, \cite{tony2011optimal}). The second problem is the separation of signals from noise variables, for which multiple testing has been used to control inflated false positive errors under high-dimensionality.
\autoref{fig:phase}, modified from \cite{cai2017large}, illustrates the theoretical demarcation of the difficulty levels of these two problems in the setting where all variables are independent. Given a sparsity level, signal intensity needs to be sufficiently large (above the undetectable region) for the signals to be detectable by a global testing procedure. To be well-separated from noise variables with negligible classification errors, signal intensity needs to be even larger (entering the classifiable region). Detailed interpretations for the classifiable boundary can be found in \cite{arias2017distribution}. Finer allocations of signals in the classifiable region corresponding to different forms of classification errors are summarized in \cite{gao2021concentration}. Other versions of \autoref{fig:phase} and related analyses under different model settings include but are not limited to \cite{donoho2004higher}, \cite{donoho2015special}, \cite{ji2012ups}, \cite{ji2014rate}, \cite{jin2017phase}, and \cite{chen2019two}.

In this paper, we aim to address the challenge of retaining weak signals that can only be detected by global testing but are not separable from noise by multiple testing, as shown in the middle white region of \autoref{fig:phase}. These weak signals are common in large-scale data analysis, including GWAS where they can be common genetic variants with small effect sizes that account for a large portion of the missing heritability. The ability to retain such signals is crucial for identifying genetic predispositions to complex traits and diseases. However, retaining such signals poses a bottleneck challenge in high-dimensional sparse inference, and new methodological developments are needed to overcome this challenge.

To retain weak signals in the middle region of \autoref{fig:phase}, we consider strategies to effectively control false negative errors. 
However, the study of false negative control has largely lagged behind in the literature.
One cannot simply switch the null and alternative hypotheses and achieve false negative control by a hypothesis testing approach because, unlike the null distribution, alternative distributions are often unknown and hard to estimate from the data. Moreover, manipulating the nominal level of a multiple testing procedure won't necessarily achieve the desired control level of false negatives, as the trade-off relationship between false negative and false positive errors depends on the signal-to-noise ratio, which is typically unknown in real-world applications.

We propose False Negative Control (FNC) screening, a data-driven method to effectively control the false negative proportion (FNP $=$ number of false negatives/number of true signals)  at a user-specified level. The FNP measure can be interpreted as the empirical Type II error of a selection rule.  
We recognize that different application scenarios may require different levels of tolerance for false negative errors, and our FNC screening procedure allows users to control the FNP at a user-specified level under conditions that permit the retention of unclassifiable signals. Furthermore, the adaptivity of our procedure to a given FNP level, such as 0.1, enables the exclusion of a certain proportion of the weakest signals, thereby reducing possibly a large number of false positives.
It is important to note that the FNP measure differs from the false non-discovery rate, which is defined as the expected proportion of false negatives among the accepted null hypotheses for a single-step multiple testing procedure \citep{genovese2002operating, sarkar2006false}. 

In the literature, false negative control methods have only appeared recently. 
For example, the AFNC method in \cite{jeng2016rare} was proposed to control the so-called signal missing rate to detect rare variants in next-generation sequencing data analysis. 
The MDR method in \cite{cai2016optimal}  uses an empirical Bayesian approach to control the mean value of FNP. Both AFNC and MDR were developed assuming independent variables. False negative control under specific dependence settings has been considered in \cite{jeng2019efficient}, where the AdSMR method was developed to control signal missing rate under block dependence. The FNC-Reg approach in \cite{JengChen2019} considers FNP control in linear regression under specific conditions on variable dependence and signal sparsity to facilitate accurate precision matrix estimation and bias mitigation in linear regression. 

In this paper, we present FNC screening for a realistic setting with arbitrary covariance dependence between variables. We propose to calibrate the overall dependence strength through a parameter that is compatible with the parameter space of the phase diagram presented in \autoref{fig:phase}. With this calibration, we can asymptotically explicate the joint effect of covariance dependence, signal sparsity, and signal intensity on the proposed method. We interpret the results using a new phase diagram, which shows the power gain of the new method in retaining weak signals.

In simulation studies, we assessed the performance of FNC screening under various commonly observed dependence structures, ranging from weak to strong according to our dependence calibration system. The results demonstrate that our method outperforms existing methods in its ability to adapt to a user-specified false negative proportion (FNP) control level.

We applied FNC screening to 145 GWAS datasets obtained from the UK Biobank, each comprising a sample size of $n=5000$. The standard GWAS procedure failed to identify any significant genetic variants due to the limited sample size. By applying FNC screening, we discovered signal variants in 99 datasets and obtained candidate sets to effectively retain these signal variants for downstream analyses. 
To validate the results, we obtained GWAS data for the same traits from a large existing study with a sample size of $500,000$. We found that a substantial portion of the variants selected by FNC screening exhibited highly significant validation p-values, suggesting that they are likely to be true signals. 

Findings from FNC screening could be valuable for a range of downstream analyses. An interesting example lies in the field of integrative genomics. The GWAS variants selected by FNC screening can be further investigated for their regulatory effects on genes by employing expression quantitative trait loci (eQTL) data. To address the possibility of false-positive variants, joint modeling approaches can be employed, incorporating linkage disequilibrium (LD) matrix information and other relevant biological data, to identify and remove false positives. This further refines the selected variant sets, enhancing the accuracy and reliability of subsequent analyses in integrative genomics and other biological contexts.

\section{Method and Theory} \label{sec:method}

We consider a two-groups model with continuous null distribution $F_0$. 
Suppose that the observed test statistics
\begin{equation} \label{def:two-groups}
X_j \sim F_0 \cdot1\{j\in I_0\}+ F_j \cdot1\{j\in I_1\}, \qquad j=1,...,m,
\end{equation}
where $I_0$ is the set of indices for noise variables, $I_1$ is the set of indices for signal variables, and $\{F_j\}_{j \in I_1}$ are the signal distributions.
All $I_0$, $I_1$, and $\{F_j\}_{j \in I_1}$ are unknown. One can perform inverse normal transformation as $Z_j = \Phi^{-1}(F_0(X_j))$, where $\Phi^{-1}$ is the inverse of the standard normal cumulative distribution. Then, we have 
\begin{equation} \label{def:model}
	Z_j \sim \Phi \cdot1\{j\in I_0\}+ G_j \cdot1\{j\in I_1\}, \qquad j=1,...,m,
\end{equation}   
where $G_j$ are the unknown signal distributions after the transformation. For presentation simplicity,  we assume that 
$G_j(t) < \Phi(t)$ for all $t \in \mathbb{R}$. i.e., signal variables tend to show larger values than noise variables. This assumption can be easily generalized to signals with two-sided effects. 

\subsection{False negative control screening}

Consider a selection rule with threshold $t$. Define the numbers of selected variables, false positives, and false negatives as
\[
\mbox{R}(t)=\sum_{j=1}^m1_{\{Z_j>t\}}, \quad \mbox{FP}(t)=\sum_{j\in I_0}1_{\{Z_j>t\}}, \quad  \mbox{FN}(t)=\sum_{j\in I_1}1_{\{Z_j\leq t\}}.
\]
The above quantities and other classification measurements including the numbers of true positives (TP) and true negatives (TN) are summarized in \autoref{tab:confusion}.

\begin{table}[!h]
	\small
	\centering
	\caption{Classification matrix of the selection rule with threshold $t$.} \label{tab:confusion}
		\begin{tabular}{l|c c|c}
			\hline
			& Not Selected & Selected & Total\\
			\hline
			Negative & TN($t$) & FP($t$) & $m-s$\\
			Positive & FN($t$) & TP($t$) & $s$\\
			\hline
			Total & $m-R(t)$ & R$(t)$ & $m$\\
			\hline
		\end{tabular} 
		
\end{table}
Note that R$(t)$ and $m$ can be directly observed. FP$(t)$, FN$(t)$, TP$(t)$, TN$(t)$, and $s (= |I_1|)$ are unknown because $I_0$ and $I_1$ are unknown. 
A generalization to two-sided signal effects can be accommodated by replacing $Z_j$ with $|Z_j|$ and only allowing $t>0$.
  
Next, define FNP with respect to $t$ as 
\begin{equation} \label{def:FNP(t)}
	\mbox{FNP}(t)=\mbox{FN}(t)/s.
\end{equation}
FNP$(t)$ may be regarded as the empirical type II error that is non-decreasing with respect to $t$. 
We aim to determine a selection threshold $\hat t$ such that the corresponding FNP($\hat t$) can be controlled at a desirable low level. 

Our new method is based on the approximation of FNP$(t)$ for a given $t$. Utilizing the fact that $s = \mbox{FN}(t)+ \mbox{TP}(t)$ and R$(t) = \mbox{FP}(t) + \mbox{TP}(t)$ as shown in Table \ref{tab:confusion}, we have
\begin{equation} \label{def:FNP}
\mbox{FNP}(t)= \mbox{FN}(t)/s= 1- (\mbox{R}(t)-\mbox{FP}(t))/s,
\end{equation}
where R$(t)$ is directly observable from the data. Because the noise distribution of $Z_j$ is $N(0,1)$ and there are $m-s$ noise variables, FP$(t)$ can be approximated by its mean value $E(\mbox{FP}(t)) = (m-s) \bar \Phi(t)$, where $\bar{\Phi}(t)=1- \Phi(t)$. For illustration purpose, we first assume that the true value of $s$ is known and define 
\begin{equation} \label{def:hatFNP}
\widehat{\mbox{FNP}}(t)= \max\{1-\mbox{R}(t)/s+ (m-s) \bar{\Phi}(t)/s, ~0\}.
\end{equation}
$\widehat{\mbox{FNP}}(t)$ with an estimated $s$ will be discussed in \autoref{sec:estimatedS}. 
Note that $\widehat{\mbox{FNP}}(t)$ is not the mean value of $\mbox{FNP}(t)$, and its construction does not require information of the signal distributions $G_j, j \in I_1$. If two-sided signal effects are under consideration, we can simply modify $\widehat{\mbox{FNP}}(t)$ by replacing $(m-s)$ with $2(m-s)$. 

Now given a user-specified control level, $\beta (>0)$, on FNP, we determine the selection threshold as 
\begin{equation} \label{def:hat_t}
\hat{t}(\beta)=\sup{\{t:\widehat{\mbox{FNP}}(t) < \beta\}},
\end{equation}
and select all the candidates with $Z_j > \hat t (\beta)$. If two-sided signal effects are considered, all candidates with $|Z_j| > \hat t (\beta)$ will be selected. We refer to this procedure as False Negative Control (FNC) screening.  

The proposed FNC screening procedure has several key features. 
(1) When all variables are ranked such that $Z_{(1)} \ge Z_{(2)} \ge \ldots \ge Z_{(m)}$,  FNC screening selects the smallest subset from the top whose estimated FNP is less than $\beta$. Such selection process can effectively retain a high proportion of true signals while excluding unnecessary noise variables. 
(2) FNC screening can be implemented as long as the marginal distribution of noise variables is known. The procedure does not require any information of the signal distributions nor dependence between the variables. 
(3) In real studies, researchers may have different tolerance levels for false negative errors. Adapting to a user-specified control level on FNP can substantially increase the applicability of the method. 
(4) From a practical standpoint, allowing for a certain percentage of FNP may avoid possibly a large number of false positives as a trade off. 

We would like to point out that although there exists a relationship between FNP$(t)$ and false discovery proportion $(\mbox{FDP}(t) = \mbox{FP}(t) / \mbox{R}(t))$ through 
\[
\mbox{FNP}(t) = 1 -{\mbox{R}(t)\over s}+ {\mbox{FDP}(t) \cdot \mbox{R}(t) \over s},
\]
consistent estimation of $\mbox{FDP}(t)$ by an estimator $\widehat{\mbox{FDP}}(t)$ does not imply consistent estimation of $\mbox{FNP}(t)$ by $1 - \mbox{R}(t)/s+ \widehat{\mbox{FDP}}(t) \cdot \mbox{R}(t) / s$. This is because the estimation error for $\mbox{FNP}(t)$ also depends on  the factor of $\mbox{R}(t)/ s$, whose magnitude varies with $t$ and can be on the order of $O(m/s)$ for $t = O(1)$.

\subsection{Theoretical properties under the Gaussian assumption} \label{sec:FNPcontrol}

In this section, we study the asymptotic behavior of the proposed FNC screening method in the setting of sparse inference where signal proportion $\pi (= s/m$) converges to zero as $m \to \infty$. Specifically, we assume
\begin{equation} \label{def:proportion}
\pi = \pi_m = m^{-\gamma}, \qquad \gamma \in (0, 1).
\end{equation}
Consequently, the number of signals $s (= m^{1-\gamma})$ increases with $m$ but of a smaller order.
Such calibration on sparsity has been widely adopted in literature, see e.g. \cite{donoho2004higher}, \cite{tony2011optimal}, \cite{arias2011global}, \cite{ji2012ups}, \cite{ji2014rate}, \cite{jin2017phase}.

In this section, we explicate the combined effects of signal sparsity, signal intensity, and dependence between variables under the assumption that
\begin{equation} \label{def:multiNorm}
(Z_1, \ldots, Z_m) \sim N_m(\mu, \mathbf{\Sigma}),
\end{equation}
where $\mu$ is a $m$-dimensional vector with $\mu_j= A_j \cdot 1\{j\in I_1\}$, $A_j > 0$, and $\mathbf{\Sigma}$ is an unknown and arbitrary correlation matrix. 

Different from the model setting of \autoref{fig:phase}, where $Z_1, \ldots, Z_m$, are independently distributed Gaussian variables, we consider arbitrarily correlated $Z_j$ and study dependence effect on the proposed method. To this end, we calibrate the dependence effect through the following procedure: 

(a) Define 
\[
\bar{\rho}=\lVert\mathbf{\Sigma}\rVert_1/m^2, \qquad \text{where~} \lVert\mathbf{\Sigma}\rVert_1 = \sum_{ij} |\sigma_{ij}|;
\]

and (b) introduce parameter $\eta$ such that
\begin{equation} \label{def:eta}
\bar \rho = \bar \rho_m = m^{-\eta} , \qquad \eta \in [0, 1].
\end{equation}
In can be seen that $\bar \rho$ summarizes the overall covariance dependence by calculating the mean absolute correlation. 
In high-dimensional data analysis with large $m$, $\bar \rho$ is often very close to zero because not every variable is correlated to all the other variables. For example, $\bar \rho$ of the $\mathbf{\Sigma}_{m \times m}$ from an autoregressive model has the order of $m^{-1}$. In order to  calibrate the dependence effect in a proper scale, we introduce the parameter $\eta$.  Clearly, $\eta$ is in a constant scale and decreases with $\bar \rho$. $\eta=0$ corresponds to the extremely dependent case where every variable is correlated to all the other variables, and, at the opposite end, $\eta=1$ corresponds to the independent case. In the literature, dependence conditions that are more general than (\ref{def:eta}) have been imposed for related inference problems such as multiple testing \citep{fan2012estimating} and exact support recovery \citep{gao2020fundamental}. The new calibration in (\ref{def:eta}) enables the following joint analysis of signal sparsity and covariance effects for FNP control.

With $\gamma$ and $\eta$ representing signal sparsity and overall strength of covariance dependence, respectively, we discover a lower bound condition on the signal intensity ($A_j$) for the success of the proposed method. The lower bound is defined as
\begin{equation} \label{def:t_min}
\mu_{min}= \min\{\mu_1, ~\mu_2\}, 
\end{equation}
where 
\[
\mu_1 = \sqrt{2\gamma \log{m}} \qquad \mbox{and} \qquad \mu_2 = \sqrt{(4\gamma-2\eta)_{+} \log{m} + 4 \log_{(3)} m}. 
\]
Note that the lower bound $\mu_{min}$ increases with $m$ at the order of $\sqrt{\log m}$. Such order is frequently required for signal intensity level in high-dimensional sparse inference (see, e.g., \cite{ingster1994minimax}, \cite{donoho2004higher}, \cite{tony2011optimal}, \cite{arias2011global}, and \cite{cai2017large}). 
The term $\log_{(3)} m (= \log \log \log m$) is a technical term of an negligible order.

The lower bound $\mu_{min}$ takes the value of either $\mu_1$ or $\mu_2$, depending on whichever is smaller under dependence. Specifically, $\mu_2<\mu_1$ when $\eta$ is large enough or, in other words, when covariance dependence is weak enough. As dependence gets stronger and $\eta$ gets smaller, $\mu_1< \mu_2$ and the lower bound equals to $\mu_1$ and stops to change with $\eta$. 

On the other hand, the lower bound $\mu_{min}$ depends on the signal sparsity parameter $\gamma$.  The term $(4 \gamma - 2 \eta)_+$ can be zero for small enough $\gamma$, which means that the required signal intensity is only $\sqrt{4 \log_{(3)}m}$ when signals are dense enough.  For example, in the special case with independent variables, we have $\eta=1$ and  $(4 \gamma - 2 \eta)_+ = 0$ for $\gamma \le 1/2$.  

The following theorem shows the asymptotic performance of FNC screening when the lower bound condition is satisfied. The proof of the Theorem is provided in Section 1 of Supplementary Material \citep{jeng2023suppl}.

\begin{theorem}\label{thm:FNPcontrol} 
	Consider model (\ref{def:multiNorm}) and a user-specified control level $\beta$ of FNP. Let $A_{min} = \min\{A_j, j\in I_1\}$ and assume $A_{min}- \mu_{min} \to \infty$ as $m\to \infty$, where $\mu_{min}$ is defined in (\ref{def:t_min}). Then the FNC screening procedure with threshold $\hat{t}(\beta)$ defined in (\ref{def:hat_t}) efficiently controls the true FNP at the level of $\beta$, i.e.,
	\begin{equation} \label{eq:FNPcontrol_1}
	P(\mbox{FNP}(\hat{t}(\beta)) \le \beta)\rightarrow 1,
	\end{equation}
	and, for any $\tilde{t} > \hat{t}(\beta)$, 
	\begin{equation} \label{eq:FNPcontrol_2}
	P\{\mbox{FNP}(\tilde{t})>\beta -\delta) \rightarrow 1 
	\end{equation}
	for arbitrarily small constant $\delta>0$. 
\end{theorem}

\autoref{thm:FNPcontrol} shows efficiency of FNC screening in selecting the smallest subset of variables to achieve FNP control at the $\beta$ level. In other words, the method also regulates unnecessary false positives to achieve the user-specified FNP control.  
The lower bound condition on $A_{min}$ explicates the joint effect of signal intensity, signal sparsity, and  covariance dependence on the FNC screening method. 
\begin{figure}[h]
	\centering
	\includegraphics[height = 6.8cm, width=7cm]{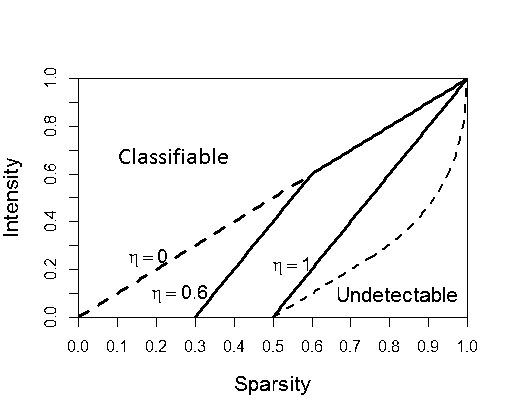}
	\vspace{-0.1in}
	\caption{Signal retainable region of FNC screening under covariance dependence. Signals in the area above a solid red line can be retained by FNC screening at a pre-specified FNP level. The area increases with $\eta$ as the dependence get weaker. } \label{fig:phase_fnp}
\end{figure}

Utilizing the calibration of covariance dependence through $\eta$, we present the results of \autoref{thm:FNPcontrol} in a new phase diagram as follows. Let $A_{min} = \sqrt{2 r \log m}$, $r > 0$. The condition on $A_{min}$ in \autoref{thm:FNPcontrol} can be transformed to $r > \min\{\gamma, ~2\gamma - \eta\}$ and demonstrated in a two-dimensional phase diagram with $\gamma$ as the x axis and $r$ as the y axis. The lower bound  $\min\{\gamma, ~2\gamma - \eta\}$  is illustrated in \autoref{fig:phase_fnp} by the solid line that moves with the dependence parameter $\eta$. Signals in the area above the solid line can be efficiently retained by FNC screening at a pre-specified level. Recall \autoref{fig:phase}, which shows that under independence, signals in the region between the two dashed lines are only detectable for their existence but not classifiable. Our result in the special case of independence ($\eta =1$) shows that many of such weak signals can be efficiently retained by the proposed method.




	\subsection{FNP control with estimated number of signals} \label{sec:estimatedS}
	
	Implementation of the proposed method requires information of the number of signals ($s$) which, in real applications, is often unknown. Existing studies for the estimation of $s$ usually assume independence among variables \citep{GW04, MR06, cai2007estimation, cai2010optimal}, and most of them are for relatively dense signals. We are interested in finding an estimator of $s$ that works in our setting with arbitrary covariance dependence. Recent study in \cite{jeng2023estimating} introduces an estimator for the signal proportion $\pi (= s/m)$ with the form $\hat \pi_{} = \max\{\hat \pi_\delta, \delta \in \Delta\}$, where $\Delta$ is a set of functions that render the most powerful consistent estimators 
	in a family of estimators under different dependence scenarios. More specifically, the family of estimators are defined as 
	\begin{equation} \label{def:MR}
		\hat{\pi}_{\delta}=\sup_{t > 0}{\frac{ R(t)/m-2\bar{\Phi}(t)-c_{m, \delta}\delta \left( t\right) }{1-2\bar{\Phi}(t)}}, 
	\end{equation}
	where $\delta(t)$ is a strictly positive function on $(0, \infty)$ and
	$c_{m, \delta}$ is the bounding sequence corresponding to $\delta(t)$. It has been discovered that  $\delta(t) = [\bar \Phi(t)]^{1/2}$ and $\delta(t) =\bar \Phi(t)$ result in powerful $\hat{\pi}_{\delta}$ under different dependence scenarios, and a new estimator is developed as 
	\begin{equation} \label{def:hat_pi}
		\hat \pi_{} = \max\{\hat \pi_{0.5}, \hat \pi_{1}\},
	\end{equation}
	where $\hat \pi_{0.5}$ denotes $\hat \pi_{\delta}$ with $\delta = [\bar \Phi(t)]^{1/2}$, and $\hat \pi_{1}$ denotes $\hat \pi_{\delta}$ with $\delta = \bar \Phi(t)$. Details about the selection of  $\delta(t)$, the construction of $c_{m, \delta}$, and the properties of  $\hat{\pi}_{\delta}$ and $\hat \pi_{}$ can be found in \cite{jeng2023estimating}. 
	
	We extend the results in \cite{jeng2023estimating} to our setting with $\gamma$ and $\eta$ representing signal sparsity and overall dependence strength, respectively, so that the conditions for the consistency of  $\hat \pi_{}$ and the conditions for the FNP control of our proposed method can be unified. Because the theoretical analysis in \cite{jeng2023estimating} is conducted on a discretized version ($\hat \pi_{\delta}^*$) of $\hat \pi_{\delta}$, which is defined by replacing $\sup_{t>0}$ in (\ref{def:MR}) with $\max_{t \in \mathbb{T}}$, where $\mathbb{T} = [\sqrt{\log \log m}, \sqrt{5 \log m}] \cap \mathbb{N}$, we present the unified conditions for the discretized version of  $\hat \pi_{}$ as well and define it as  $\hat \pi_{}^* = \max\{\hat \pi^*_{0.5}, \hat \pi^*_{1}\}$.

	\begin{proposition} \label{prop:estimated_s}
		Assume the same conditions as in \autoref{thm:FNPcontrol}.
		Then, for any constant $\epsilon>0$,  
		\begin{equation} \label{eq:hat_pi}
			P((1-\epsilon) < \hat \pi_{}^*/\pi  < 1)\rightarrow 1.
		\end{equation} 
		Replace $s$ in (\ref{def:hat_t}) with the estimator $\hat s = m \cdot \hat \pi_{}^*$ and denote the selection threshold as $\hat t_{\hat s}(\beta)$. We have
		\begin{equation} \label{eq:FNPcontrol_1s}
			P(\mbox{FNP}(\hat{t}_{\hat s}(\beta)) \le \beta)\rightarrow 1
		\end{equation}
		and, for any threshold $\tilde{t} > \hat{t}_{\hat s}(\beta)$, 
		\begin{equation} \label{eq:FNPcontrol_2s}
			P\{\mbox{FNP}(\Tilde{t})>\beta -\delta) \rightarrow 1 
		\end{equation}
		for arbitrarily small constant $\delta>0$. 	
	\end{proposition}
	
	Proposition \autoref{prop:estimated_s} shows that the same set of conditions are sufficient for both the consistency of $\hat \pi_{}^*$ and the efficient FNP control of FNC screening implemented with $\hat \pi_{}^*$. This result can also be interpreted by the phase diagram in \autoref{fig:phase_fnp}, where the solid red line that moves with the dependence level $\eta$ now indicates the required signal intensity for both $s$ estimation and FNP control. In other words, for signals above the solid red line, although they may not be indivily identifiable, 
	we can still perform meaningful inference by not only estimating their total numbers but also effectively retaining them at a target FNP level.   
	
	Generally speaking, if we want to study the power of some method, a condition on the signal intensity is unavoidable. When signal intensity is too low, no methods can even detect their existence as illustrated in \autoref{fig:phase}, let along the more challenging tasks of FNP control. 
	When the condition on signal intensity does not hold, the estimator $\hat \pi_{}$ tends to underestimate  because of its lower bound property \citep{jeng2023estimating}. Consequently, FNC screening implemented with $\hat s = m \cdot \hat{\pi}_{}$ tends to have a threshold higher than the ideal threshold that achieves the target level of $\beta$, which results in conservative variable selection with less false positives but inflated FNP. This tendency is observed in simulation examples with low signal-to-noise ratio in Section \ref{sec:simulation_estS}.

We note that FNC screening is closely related to the FNC-Reg approach in \cite{JengChen2019}. However, these two methods are developed for different models (two-groups model vs. linear regression model) and are studied under very different conditions. Particularly, the success of FNC-reg requires specific conditions on variable dependence and signal sparsity to facilitate accurate precision matrix estimation and bias mitigation in linear regression. Moreover, FNC-reg uses a different estimator for the number of signals, whose consistency under arbitrary covariance dependence of the two-groups model is unclear.

\subsection{Computational algorithms}

We provide algorithms to calculate the bounding sequence $c_{m, \delta}$, to derive the proportion estimator $\hat \pi$ in (\ref{def:MR}), and to select variables by FNC screening.

\begin{algorithm}[!h]
	\caption{Bounding sequence $c_{m, \delta}$} \label{alg_1}	
	\begin{algorithmic}[1]
		\Statex {\bf Input:} $N$ sets of $p$-values generated by the joint null distribution of $\{z_1, \ldots z_m\}$
		\Statex {\bf Output:} bounding sequences $c_{m, 0.5}$ and $c_{m, 1}$ 
		\State {\bf for} $a=1, 2\ldots, N$ {\bf do} 
		\Statex Rank the $a$-th set of p-values such that $p_{a,(1)} < p_{a,(2)} < \ldots < p_{a,(m)}$
		\Statex Compute 
		\[
		V_{a, 0.5} = \max_{1 < j < m} {|j / m - p_{a, (j)} | \over \sqrt{p_{a,(j)}}} \qquad \text{and} \qquad V_{a, 1} = \max_{1 < j < m} {|j / m - p_{a, (j)} | \over p_{a,(j)}}  
		\]
		\State {\bf end for}
		\State Compute $c_{m, 0.5}$ as the $(1-1/\sqrt{\log m})$-th quantile of the empirical distribution of $V_{a, 0.5}, a = 1, \ldots N$ and compute $c_{m, 1}$ as the $(1-1/\sqrt{\log m})$-th quantile of the empirical distribution of $V_{a, 1}, a = 1, \ldots N$.
	\end{algorithmic}
\end{algorithm}	

The computation of  $c_{m, \delta}$ in \autoref{alg_1} requires input of $p$-values generated by the joint null distribution. When the joint null distribution is unknown in  real applications, we can often simulate the null $p$-values non-parametrically. 
For example, when$\{z_1, \ldots z_m\}$ are a set of test statistics for associations between a set of explanatory variables and a response variable, we may randomly shuffle only the sample of the response variable to remove potential associations with the explanatory variables and then calculate the test statistics. More details for such permutation approaches can be found in \cite{westfall1993resampling}. We generate $N$ sets of null $p$-values, where $N$ is a predetermined large number, such as 1000.

\begin{algorithm}[]
	\caption{Signal Proportion Estimator}\label{alg}	
	\begin{algorithmic}[1]
		\Statex {\bf Input:} $p$-values of the observed test statistics and bounding sequences $c_{m, 0.5}$ and $c_{m, 1}$
		\Statex {\bf Output:} a proportion estimate $\hat \pi$
		\State Rank the variables by their $p$-values so that $p_{(1)} < p_{(2)} < \ldots < p_{(m)}$	
		\State Compute 
		\[
		\hat{\pi}_{0.5}= \max_{1 < j < m}\frac{ j/m-p_{(j)}-c_{m,0.5} \cdot \sqrt{p_{(j)}}} {1-p_{(j)}} \qquad \mbox{and} \qquad \hat{\pi}_{1}= \max_{1 < j < m}\frac{ j/m-p_{(j)} -c_{m,1} \cdot p_{(j)}} {1-p_{(j)}}
		\]
		\State Obtain $\hat \pi = \max\{\hat{\pi}_{0.5}, \hat{\pi}_{1}\}$
	\end{algorithmic}
\end{algorithm}

\begin{algorithm}[]
	\caption{FNC Screening}\label{alg}	
	\begin{algorithmic}[1]	
		\Statex {\bf Input:} $p$-values of the observed test statistics, a user-specified $\beta$, and a proportion estimate $\hat \pi$
		\Statex {\bf Output:} a set of selected variables		
		\State Rank the variables by their $p$-values so that $p_{(1)} < p_{(2)} < \ldots < p_{(m)}$	
		\State Compute $\hat s = \hat \pi m$
		\State For $j=1, 2, \ldots, m$, compute
		\[
		\widehat{FNP}_j= \max\{1 - j/\hat s + (m-\hat s) p_{(j)}/\hat s, ~0\}
		\]		
		\State Let $k=1$
		\Statex {\bf while}  $ \widehat{FNP}_k 
		\ge \beta$ {\bf do}  $k = k+1$
		\Statex {\bf end while} 	
		\State Obtain variables ranked at $1, \ldots, k-1$		
	\end{algorithmic}
\end{algorithm}

\section{Simulation}  \label{sec:simulation}
	
We provide simulation examples to demonstrate the finite-sample performance of the proposed FNC screening procedure. In these examples, 
a series of variables are generated as 
	\begin{equation} \label{eq:Z}
	(Z_1, \ldots, Z_m)^T \sim N((\mu_1, \ldots, \mu_m)^T, \mathbf{\Sigma}),
	\end{equation}
	where $\mu_j = A \cdot 1\{j \in I_1\}$, and $I_1$ is a set of indices randomly sampled from $\{1,..., m\}$ with cardinality $s = |I_1| = m^{1-\gamma}$.  We consider three commonly observed dependence structures:
	\begin{itemize}
		\item Model 1 [Autoregressive]: $\mathbf{\Sigma}=(\sigma^{(1)}_{ij})$, where $\sigma^{(1)}_{ij}=\lambda^{|i-j|}$ for $1\leq i,j\leq m$.
		\item Model 2 [Block dependence]: $\mathbf{\Sigma}=\mathbf{I}_{m/k}\otimes\mathbf{D}$, where $\mathbf{D}$ is a $k\times k$ matrix with diagonal entries 1 and off-diagonal entries $r$.
		\item Model 3 [Factor model]: $\mathbf{\Sigma}=(\sigma^{(3)}_{ij})$, where $\sigma^{(3)}_{ij}=V_{ij}/\sqrt{V_{ii}V_{jj}}$ for $1\leq i,j\leq m$, $\mathbf{V}=\tau\mathbf{h}\mathbf{h}^T+\mathbf{I}_m$ with $\tau \in (0, 1)$ and $\mathbf{h}\sim N(\mathbf{0}, \mathbf{I}_m)$.
	\end{itemize}

For the parameters in Model 1-3, we set $\lambda=0.2, k=40, r=0.5$, and $\tau=0.5$. The corresponding values of $\eta$, calculated by (\ref{def:eta}) with $m=2000$ or $10,000$, are presented in \autoref{tab:eta_mu}. It shows that dependence is very weak in Model 1 [Autoregressive] with $\eta$ close to 1, moderately weak in Model 2 [Block dependence] with $\eta$ between $0.6$ and $0.7$, and very strong in Model 3 [Factor model] with $\eta$ between $0.1$ and $0.3$. 

Recall the lower bound condition on the signal intensity in (\ref{def:t_min}): $\mu_{min} = \min\{\mu_1, \mu_2\}$, where $\mu_1$ depends on signal sparsity through $\gamma$,  $\mu_2$ depends on both signal sparsity and dependence through $\gamma$ and $\eta$, respectively.  We present the values of $\mu_1$, $\mu_2$, and $\mu_{min}$ under different dependence models in \autoref{tab:eta_mu}.

	\begin{table}[!h]
	\caption{Values of the dependence parameter $\eta$ and the values of $\mu_1$, $\mu_2$, and $\mu_{min}$ in simulation examples with $\gamma = 0.3$.} \label{tab:eta_mu}
	\centering
	\setlength{\tabcolsep}{1.5mm}{
		\begin{tabular}{ll| c c c c }
			& & $\eta$ & $\mu_1$ & $\mu_2$ & $\mu_{min}$   \\
			\hline
			$m=2,000$ & Autoregressive & 0.95 & 2.14 & 1.68 & 1.68  \\
			& Block dependence & 0.60 & 2.14 & 1.68 & 1.68 \\
			& Factor model & 0.22 & 2.14 & 2.94 & 2.14 \\
		    \hline
			$m=10,000$ & Autoregressive & 0.96 & 2.35 & 1.79 & 1.79 \\
			& Block dependence & 0.67 & 2.35 & 1.79 & 1.79 \\
			& Factor model & 0.18 & 2.35 & 3.29 & 2.35 \\
			\hline
	\end{tabular}}
\end{table}

\subsection{Comparison with multiple testing methods}	\label{sec:sim_multiple}

We compare the proposed method with the classical BH-FDR procedure \citep{benjamini1995}. Specifically, we show that the FNP control property of our method cannot be achieved by manipulating the nominal level of BH-FDR. 

As shown in \autoref{tab:eta_mu}, when $m=2,000$ and $\gamma=0.3$, $\mu_{min}$ of the lower bound condition is in the range of $1.68 - 2.14$ for the three dependence models. In this section, we set the signal intensity $A = 2$ and 3, apply FNC screening method with the $s$ known to us, and present the realized FNP of FNC screening from 100 replications. The results are compared with those of BH-FDR in \autoref{tab:compare}. 
We also compare the realized false discovery proportion (FDP) of the two methods. Both methods are applied with varying nominal levels.

	\begin{table}[!h]
		\caption{Mean values and standard deviations (in brackets) of the realized FNP and FDP of the proposed FNC screening (FNCS) method and the existing BH-FDR procedure. This set of examples have $m=2000$ and $\gamma=0.3$.  
		} \label{tab:compare}
		\centering
		\setlength{\tabcolsep}{1.5mm}{
			\begin{tabular}{ll l| c c }
				& & & FNP & FDP   \\
				\hline
				A=2 & Autoregressive & FNCS(${\bf \beta= 0.2}$) & {\bf 0.201} (0.069) & 0.576  (0.073) \\
				& & FNCS($\beta={\bf 0.1}$) & {\bf 0.114} (0.069)  & 0.688 (0.085) \\
				& & BH-FDR($\alpha={\bf 0.05}$) & 0.889 (0.035)  & {\bf 0.040} (0.043) \\
				& & BH-FDR($\alpha={\bf 0.2}$) & 0.630 (0.050)  & {\bf 0.181} (0.044) \\
				\cline{2-5}
				& Block dependence & FNCS($\beta={\bf 0.2}$) & {\bf 0.193} (0.132)  & 0.607 (0.161) \\
				& & FNCS($\beta={\bf 0.1}$) & {\bf 0.132} (0.116) & 0.685 (0.152) \\
				& & BH-FDR($\alpha={\bf 0.05}$) & 0.894 (0.058)  & {\bf 0.055} (0.070) \\
				& & BH-FDR($\alpha={\bf 0.2}$) & 0.647 (0.097)  & {\bf 0.182} (0.093) \\
				\cline{2-5}
				& Factor model & FNCS($\beta={\bf 0.2}$) & {\bf 0.188} (0.169) &  0.642 (0.107) \\
				& & FNCS($\beta={\bf 0.1}$) & {\bf 0.156} (0.159) & 0.693 (0.098) \\
				& & BH-FDR($\alpha={\bf 0.05}$) & 0.886 (0.088) & {\bf 0.032} (0.061) \\
				& & BH-FDR($\alpha={\bf 0.2}$) & 0.659 (0.088)  & {\bf 0.160} (0.123) \\
				\hline
				A=3 & Autoregressive & FNCS(${\bf \beta= 0.2}$) & {\bf 0.198} (0.023) &  0.149 (0.035) \\
				& & FNCS($\beta={\bf 0.1}$) & {\bf 0.101} (0.037)  & 0.307 (0.084) \\
				& & BH-FDR($\alpha={\bf 0.05}$) & 0.378 (0.040)  & {\bf 0.044} (0.018) \\
				& & BH-FDR($\alpha={\bf 0.2}$) & 0.166 (0.028) & {\bf 0.183} (0.028) \\
				\cline{2-5} 		
				& Block dependence & FNCS($\beta={\bf 0.2}$) & {\bf 0.170} (0.086) & 0.259 (0.247) \\
				& & FNCS($\beta={\bf 0.1}$) & {\bf 0.089} (0.074) & 0.444 (0.268) \\
				& & BH-FDR($\alpha={\bf 0.05}$) & 0.387 (0.071) & {\bf 0.047} (0.036) \\
				& & BH-FDR($\alpha={\bf 0.2}$) & 0.169 (0.049)  & {\bf 0.180} (0.070) \\
				\cline{2-5}
				& Factor model & FNCS($\beta={\bf 0.2}$) & {\bf 0.160} (0.099) &  0.262 (0.215) \\
				& & FNCS($\beta={\bf 0.1}$) & {\bf 0.084} (0.104) &  0.533 (0.227) \\
				& & BH-FDR($\alpha={\bf 0.05}$) & 0.400 (0.043) & {\bf 0.041} (0.049) \\
				& & BH-FDR($\alpha={\bf 0.2}$) & 0.176 (0.049)  & {\bf 0.170} (0.113) \\
				\hline
			\end{tabular}}
		\end{table}
		
		It can be seen that the mean values of FNP for FNC screening are generally closer to their corresponding nominal levels ($\beta=0.2$ and $0.1$) with larger $A$ values. 
		As dependence gets stronger from Model 1 [Autoregression] to Model 2 [Block dependence], to Model 3 [Factor model], not only the differences between the realized FNP and the nominal levels increases, but the standard deviations of the realized FNP increases as well. In other words, effective control of FNP by FNC screening becomes harder under stronger dependence. These observations agree with the theoretical results in Section \ref{sec:FNPcontrol}. 
		
		On the other hand, \autoref{tab:compare} shows that the mean values of the realized FDP of BH-FDR are fairly close to their corresponding nominal levels of $\alpha$. 
		However, the mean values of FNP for BH-FDR are generally larger than those of FNC screening. It is true that one can increase the nominal level of BH-FDR to reduce FNP, but it is unclear by how much one should increase $\alpha$ to achieve a target level of FNP. For example, suppose we have a target $\beta = 0.2$. It seems that when $A=3$, BH-FDR with $\alpha=0.2$ has the mean values of FNP in the range of $0.166-0.176$, which are close to the target $\beta$. 
		However, when $A=2$, BH-FDR with $\alpha=0.2$ has mean values of FNP in the range of $0.63 - 0.66$, which are much bigger than $\beta=0.2$.
		Analogous results for $m=10,000$ and $\gamma=0.3$ are presented in Section 2 of Supplementary Material \citep{jeng2023suppl}.
		
		These results demonstrate the fundamental differences between the existing multiple testing procedures and the proposed FNC screening method as they serve for very different purposes. The proposed method aims to efficiently retain signals at a target FNP level, for which the multiple testing methods cannot achieve. On the other hand, the new method pays the price of having a higher FDP when signals are relatively weak.

	\subsection{Comparison with other false negative control methods} \label{sec:simulation_estS}
	In this section,  we compare the empirical performances of FNC screening with existing methods that are designed to regulate false negatives. Among the existing methods discussed in Section \ref{sec:introduction}, AFNC and MDR are more comparable to FNC screening because they all consider the two-groups model as in (\ref{def:two-groups}) and require the input of a user-specified control level. Moreover, they all require an estimate for the number of signals. For a fair comparison, we implement the same estimator $\hat s = m \cdot \hat \pi$, where $\hat \pi$ is defined in (\ref{def:hat_pi}), to these methods.

	
	
	The performances of the methods are evaluated by three measures. The first two measures, FNP and FDP are the same as in \autoref{tab:compare}.  The third measure is the Fowlkes-Mallows index \citep{fowlkes1983method, halkidi2001clustering}, which summarizes the measures of FNP and FDP by calculating the geometric mean of ($1-$FNP) and ($1-$FDP), i.e., 
	\[
	\mbox{FM-index} = \sqrt{(1-\mbox{FNP}) \times (1-\mbox{FDP})}.
	\]   
	Higher values of the FM-index indicate better classification results. The FM-index is a sensible summary measure in high-dimensional settings with sparse alternative cases because the scale of FDP is more comparable to that of FNP than the classical false positive proportion (FPP = number of false positives/number of null cases). Because the methods presented in this section all focus on false negative control, it is appropriate to use FM-index to compare their efficiency.  We also report the number of selected variables (\#R) by each method. 
	
	In this section, variables are generated by (\ref{eq:Z}) with $m=2,000$ and a covariance matrix that has 20 diagonal blocks with block sizes randomly generated from 10 to 100. The non-zero off-diagonal correlations are set at 0.5. The dependence parameter $\eta$ varies from sample to sample due to the random block size.

	In the first set of examples, signal sparsity is fixed with $\gamma = 0.3$, and signal intensity (A) increases from $3$ to $5$.  First, 
	because the estimator $\hat s$ underestimates the true $s$ when signal intensity is not strong enough, FNC screening implemented with $\hat s$ selects less variables than actually needed to reach the nominal level of $\beta$. These result in inflated realized FNP as seen in \autoref{tab:compare_mu}. 
	Further, as $A$ increases, the mean value of FNP of FNC screening gets closer to the nominal level of $\beta$, which agrees with the claims in (\ref{eq:FNPcontrol_1s}) and (\ref{eq:FNPcontrol_2s}) in Proposition \ref{prop:estimated_s}.  This tendency, however, is not observed for the other two methods presented in \autoref{tab:compare_mu}. Among the three methods, FNC screening selects the fewest number of variables and generally outperforms the other two methods in terms of the FM-index, particularly when the signal intensity increases.  
	
	\begin{table}[t]
		\caption{Effect of signal intensity on different FN control methods. Mean values and standard deviations (in brackets) of the realized FNP, FDP, the FM-index, and the number of selected variables (R) are presented for the proposed FNC screening method and two existing methods, AFNC and MDR, from 100 replications. } \label{tab:compare_mu}
		\centering
		\setlength{\tabcolsep}{1.5mm}{
			\begin{tabular}{l l| c c c c}
				& &  FNP & FDP & FM-index & R \\
				\hline 
				$A=3$  & FNCS($\beta={\bf 0.1}$) & {\bf 0.28} (0.12) &  0.13 (0.14) & {\bf 0.78} (0.04) & 181 (77) \\
				& AFNC($\beta={\bf 0.1}$) &  0.19 (0.11)  &  0.22 (0.18) & 0.78 (0.07) & 239 (118) \\
				& MDR($\beta={\bf 0.1}$) &  0.09 (0.08) & 0.49 (0.26) &  0.65 (0.15) & 498 (264) \\
				\hline
				$A=4$ & FNCS($\beta={\bf 0.1}$)  & {\bf 0.16} (0.07) & 0.06 (0.12) &  {\bf 0.88} (0.05) & 192 (62) \\
				& AFNC($\beta={\bf 0.1}$) & 0.07 (0.04) &  0.14 (0.18) & 0.89 (0.10) & 242 (103) \\
				& MDR($\beta={\bf 0.1}$) &  0.06 (0.05) & 0.37 (0.32) & 0.74 (0.18) & 418 (263) \\
				\hline
				$A=5$ & FNCS($\beta={\bf 0.1}$) &  {\bf 0.10} (0.04)  & 0.04 (0.12)  & {\bf 0.93} (0.06) & 201 (57) \\
				& AFNC($\beta={\bf 0.1}$) & 0.01 (0.01)  & 0.36 (0.35) & 0.75 (0.26) & 639 (657) \\
				& MDR($\beta={\bf 0.1}$) & 0.04 (0.04) & 0.36 (0.33)  & 0.75 (0.19) & 361 (324) \\
				\hline
			\end{tabular}}
		\end{table}
		
		In the second set of examples, signal intensity is fixed at $5$ and signal sparsity $\gamma$ increases from 0.3 to 0.5, so that the number of signals deceases from 205 to 45. The nominal level of $\beta$ also varies. Results summarized in \autoref{tab:compare_beta} show that FNC screening continues to outperform the other two methods in adapting to different nominal levels and incurring less false positives. Its advantage seems to be more prominent when signals get more sparse.  
		
		

		\begin{table}[h]
			\caption{Effect of nominal level and signal sparsity on different FN control methods. Same notations as in \autoref{tab:compare_mu} are used. }\label{tab:compare_beta}
			\centering
			\setlength{\tabcolsep}{1.5mm}{
				\begin{tabular}{l l|cccc}
					&  & FNP & FDP & FM-index & R\\
					\hline 
					$\gamma=0.3$   & FNCS($\beta={\bf 0.1}$) &  {\bf 0.10} (0.04)  & 0.04 (0.12)  & {\bf 0.93} (0.06) & 201 (57) \\
					& AFNC($\beta={\bf 0.1}$)  & 0.01 (0.01)  & 0.36 (0.35) & 0.75 (0.26) & 639 (657)\\
					& MDR($\beta={\bf 0.1}$) & 0.04 (0.04) & 0.36 (0.33)  & 0.75 (0.19) & 361 (324) \\
					\cline{2-6}
					& FNCS($\beta={\bf 0.2}$)  & {\bf 0.19} (0.06) & 0.02 (0.08) & {\bf 0.89} (0.03) & 174 (39) \\
					& AFNC($\beta={\bf 0.2}$)  & 0.01 (0.01) & 0.21 (0.29) & 0.86 (0.20) & 400 (452)\\
					& MDR($\beta={\bf 0.2}$)  & 0.08 (0.09) & 0.26 (0.28) & 0.80 (0.14) & 244 (254)\\
					\hline
					$\gamma=0.5$ & FNCS($\beta={\bf 0.1}$) & {\bf 0.09} (0.05) & 0.15 (0.29) & {\bf 0.85} (0.18) & 76 (93)\\
					& AFNC($\beta={\bf 0.1}$) & 0.00 (0.01) & 0.73 (0.28) & 0.44 (0.27) & 789 (820)\\
					& MDR($\beta={\bf 0.1}$)  & 0.03 (0.05) & 0.56 (0.44) & 0.54 (0.34) & 445 (395)\\
					\cline{2-6}
					& FNCS($\beta={\bf 0.2}$)  & {\bf 0.16} (0.09) & 0.12 (0.27) & {\bf 0.83} (0.15) & 65 (76) \\
					& AFNC($\beta={\bf 0.2}$)  & 0.01 (0.02) & 0.60 (0.36) & 0.55 (0.31) & 572 (735) \\
					& MDR($\beta={\bf 0.2}$)  & 0.07 (0.10) & 0.54 (0.45) & 0.54 (0.32) & 378 (350)	\\
					\hline
				\end{tabular}}
			\end{table}

\section{Application in GWAS} \label{sec:GWAS}

The achievements of GWAS have been witnessed in genetic dissection of various complex traits and diseases, often by utilizing very large sample sizes \citep{Hu2021RBC, Mikhaylova2021WBC, Zhao2022NG, Mahajan2022T2D}. Nevertheless, GWAS of many diseases remain underpowered, particularly for non-European ancestries \citep{Sun2022UKBminority, Liu2022ELGAN}. In this section, we demonstrate how FNC screening can be applied to retain true GWAS signals for downstream analyses when sample sizes are limited.

To mimic underpowered GWAS, we randomly selected a limited number of European ancestry individuals from the UK Biobank as our study sample. We included a total of 29 blood cell traits (the same traits included in \cite{Vuckovic2020}) and randomly selected 5 genome regions (each 3-4MB in size), resulting in $29 \times 5 (=145)$ trait-region combinations/datasets, each with the sample size of $n=5000$.

For each dataset, we first applied standard GWAS with Bonferroni correction. The GWAS summary statistics were obtained using the {\it regenie} software \citep{REGENIE} with linear mixed models. Covariates, including age, sex, the first 10 genetic PCs, genotype array, and recruitment centers, were adjusted as described in \citep{Sun2022UKBminority}. The Bonferroni correction was conducted at the level of $\alpha/m_k$, where $\alpha$ is the family-wise error rate, set at $0.05$, and $m_k$ is the dependence-adjusted number of tests for the $k$th dataset. Specifically, we utilized the linkage disequilibrium (LD) information from TOP-LD \citep{Huang2022TOPLD} Europeans as a reference to approximate the number of independent tests to determine $m_k$. For all 145 datasets, no significant variants were identified by the standard GWAS approach due to the limited sample size.

In order to effectively retain signal variants for downstream analyses, we applied FNC screening and were able to discover signal variants in 99 datasets. \autoref{tab:data_compare} presents the number of variants selected by FNC screening across the 145 datasets. The results are compared to those obtained using the MDR method. While MDR often selects the majority of variants in a dataset, FNC screening is shown to be much more selective. 
 
\begin{table}[h]
	\caption{Five-number summary for the numbers of variants before and after screening by different methods across 145 working datasets.}\label{tab:data_compare}
	\centering
	\setlength{\tabcolsep}{1.5mm}{
		\begin{tabular}{l|ccccc}
			& Min & Q1 & Q2 & Q3 & Max\\
			\hline 
			Before screening   & 12939 & 14116 &  14734 &    15150 &  15636 \\
			MDR($\beta=0.1$) & 0  & 0 &  11798  & 13968  & 15422  \\
			FNCS($\beta=0.1$) & 0 & 0  & 644  & 2213 &  5546  \\
			FNCS($\beta=0.2$) & 0 & 0  & 560  & 2015 &  5333  \\
			\hline
	\end{tabular}}
\end{table}

\begin{figure}[htbp]
	\centering
	\includegraphics[width=15cm, height=9cm]{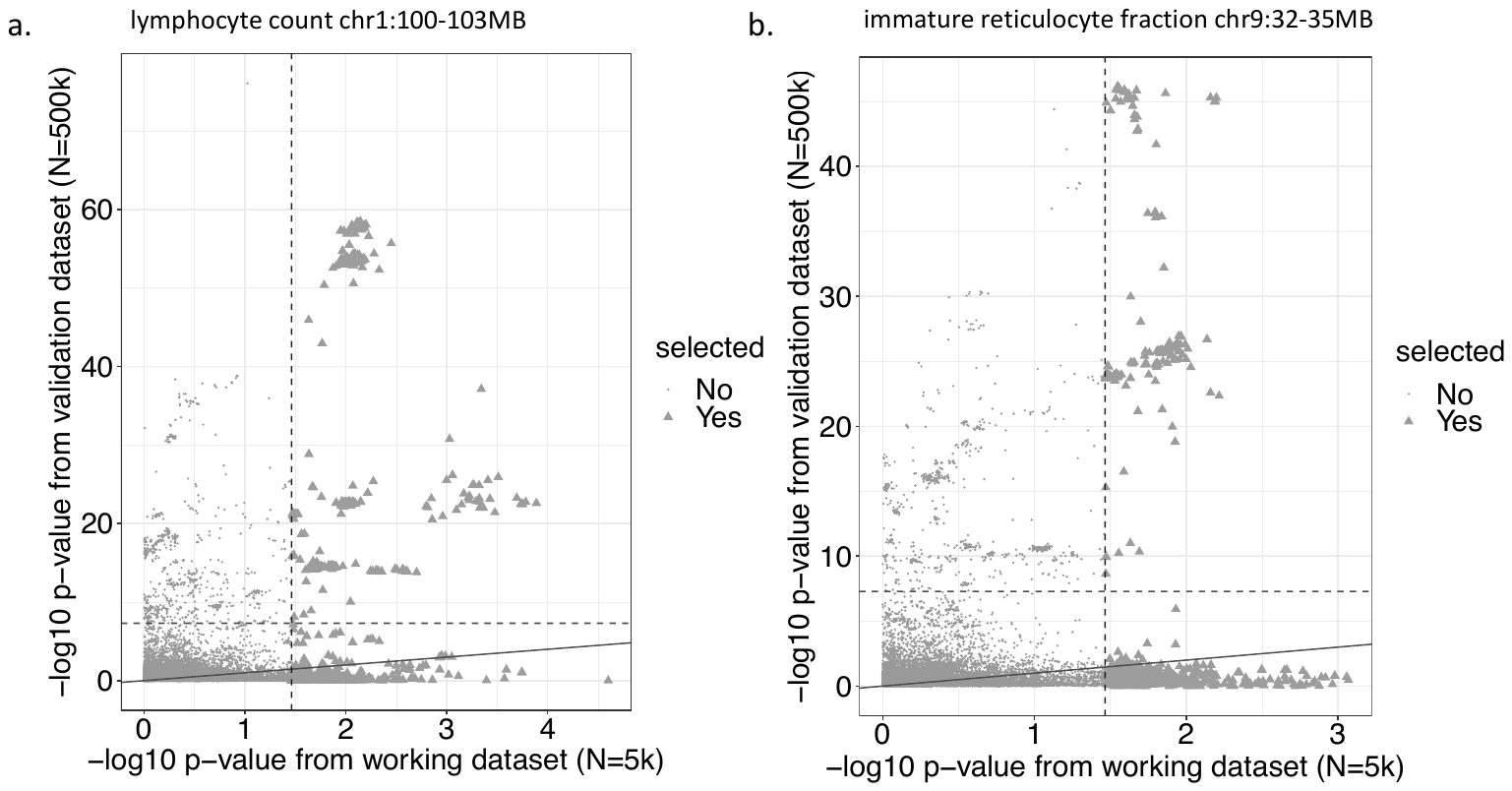}
	\vspace{-1.4in}
	\caption{Examples of the outcomes of FNC screening ($\beta = 0.1$). Panel a is for lymphocyte count on chr 1: 100-103MB, and Panel b is for immature reticulocyte fraction on chr 9: 32-35MB. Each dot in the plots represents a genetic variant, and the variants selected by FNC screening are indicated by large triangles. The $-\log_{10} p$-values for all variants, calculated based on the 5000 working sample, are compared to their $-\log_{10} p$-values calculated using the 500k validation sample. The dashed vertical line denotes the $p$-value threshold of FNC screening for the 5000 working sample, while the dashed horizontal line indicates the Bonferroni threshold for the 500k validation sample. The solid slash represents the $y=x$ 45-degree line. } \label{fig:GWAS_results}
\end{figure}

To justify the findings of FNC screening, we obtained GWAS summary statistics for the same 145 trait-region combinations based on 500,000 UK Biobank European sample from \cite{Vuckovic2020}, one of the largest GWAS for blood cell traits. We found that among the 99 datasets where signal variants were discovered by FNC screening, 33 datasets contain variants that survived Bonferroni correction in the large GWAS. \autoref{fig:GWAS_results} illustrates two such examples, where $p$-values for all the variants calculated based on the 5000 working sample are compared with the $p$-values calculated based on the 500k validation sample. Data points of the variants selected by FNC screening based on the 5000 working sample are highlighted by large triangles. It is apparent that the general pattern of the $p$-values does not exhibit good consistency across the two datasets, which is expected due to the limited sample size of the working data. However, it is noteworthy that a substantial portion of the highlighted variants have highly significant validation $p$-values, suggesting that they are likely to be true signals.  For instance, in panel a, 913 variants are selected by FNC screening, and $41.5\%$ of them have validation $p$-values less than $5 \times 10^{-8}$, the Bonferroni threshold. Similarly, in panel b, out of the 828 variants selected by FNC screening, $27.2\%$ have validation $p$-values below the Bonferroni threshold.

As FNC screening essentially benefits from pooling information from all signals to effectively retain them, its power gain is more prominent when there are a relatively large number of weak signals. This trend becomes evident when analyzing the 145 working datasets. For instance, the results presented in panel a and b of \autoref{fig:GWAS_results} have $\hat \pi = 0.028$ and 0.024, respectively, corresponding to 438 and 364 signal variants. On the other hand,  for datasets with very small signal proportion estimates, the power gain of FNC screening becomes less clear. 
Among the 99 datasets where signal variants were discovered by FNC screening, 28 had an estimated signal proportion ($\hat \pi$) less than 0.01. Out of the 28 datasets, only 3 contained variants with validation $p$-values below the Bonferroni threshold. In contrast, out of the 71 datasets that had $\hat \pi$ values greater than 0.01, 30 had significant variants in validation. In the future, it would be interesting to develop guidelines that can aid in the interpretation of results based on estimated signal proportions.	

FNC screening could be a valuable tool in integrative genomics, where the analysis of multi-omics data is impeded by the limited availability of sample sizes. Through FNC screening, GWAS signals can be effectively retained for downstream analyses. For example, variants selected by FNC screening may be further investigated for their regulatory effects on genes using eQTL data. To address the possibility of false positive variants, joint modeling approaches, incorporating LD matrix information and other relevant biological data, can be employed to identify and remove false positives. This further refines the selected variant sets, enhancing the accuracy and reliability of subsequent analyses in integrative genomics and other biological contexts.

\section{Conclusion and Discussion} \label{sec:conclusion}

Motivated by the challenge of detecting weak signals in underpowered GWASs, we develop a new analytic framework for inferring signals that are detectable yet unidentifiable. Additionally, we propose the FNC screening method as an efficient way to control the FNP at a user-specified level.

FNC screening is developed under arbitrary covariance dependence calibrated through a dependence parameter whose scale is compatible with the existing phase diagram in high-dimensional sparse inference. Utilizing the new calibration, we are able to explicate the joint effects of covariance dependence, signal sparsity, and signal intensity and interpret the results through a new phase digram to gain important insight. The implementation of FNC screening does not require any information of the dependence between variables, and the method is shown to be most powerful under independence. We note that if some dependence-related data features are known a priori, a more powerful method could be developed by leveraging the dependence information, resulting in larger signal retainable regions under dependence.

We demonstrate the finite sample performance of FNC screening in simulations with different dependence structures. The results show that FNC screening effectively retains signals at a target FNP level, which cannot be achieved by multiple testing methods. Additionally, FNC screening outperforms other false negative control methods in adapting to different nominal levels and incurring fewer false positives under dependence.

In a real-world application, we employed FNC screening to retain signal variants in GWAS when the standard GWAS procedure lacked power. Findings from FNC screening could be valuable for a range of downstream analyses.  
Overall, the proposed FNC screening method is a valuable tool for situations where false negative results significantly impede scientific investigations. Although such scenarios are widespread, there has been limited development in methodological approaches to address them. Our study presents an insightful approach to bridge this crucial methodological gap.

\begin{acks}[Acknowledgments]
The authors would like to thank the anonymous referees, an Associate
Editor and the Editor for their constructive comments that improved the
quality of this paper. This study has been conducted using the UK Biobank Resource under Application Number 25953. 
\end{acks}				

\begin{funding}
Research of Dr. Li is partially supported by NIH grant R01HL146500.
\end{funding}

\begin{supplement}
	\stitle{Theoretical proofs}
	\sdescription{Proofs for Theorem \ref{thm:FNPcontrol} and Proposition \ref{prop:estimated_s}.}
\end{supplement}

\begin{supplement}
	\stitle{Additional simulation results}
	\sdescription{Simulation results with $m=10,000$.}
\end{supplement}

\begin{supplement}
	\stitle{Programming code}
	\sdescription{R code that generates the results in Section \ref{sec:simulation}.}
\end{supplement}

\bibliographystyle{imsart-nameyear} 
\bibliography{FNP_reference}       

\newpage

\end{document}